\documentstyle[12pt]{article}

\def\Journal#1#2#3#4{#1 {\bf #2} (#4) #3}
\def\CQG#1#2#3{\Journal{Class. Quant. Grav.}{#1}{#2}{#3}}

\def\NPB#1#2#3{\Journal{Nucl. Phys.}{B#1}{#2}{#3}}
\def\PLB#1#2#3{\Journal{Phys. Lett.}{B#1}{#2}{#3}}
\def\PRL#1#2#3{\Journal{Phys. Rev. Lett.}{#1}{#2}{#3}}
\def\PRD#1#2#3{\Journal{Phys. Rev.}{D#1}{#2}{#3}}

\def\be{\begin{equation}}
\def\ee{\end{equation}}
\def\bea{\begin{eqnarray}}
\def\eea{\end{eqnarray}}
\def\bed{\begin{displaymath}}
\def\eed{\end{displaymath}}
\def\nn{\nonumber}

\def\a{\alpha}
\def\b{\beta}

\def\de{\delta}
\def\e{\epsilon}           

\def\g{\gamma}

\def\j{\psi}
\def\k{\kappa}
\def\l{\lambda}

\def\o{\omega}
\def\r{\rho}                      
\def\s{\sigma}                    

\def\J{\Psi}

\def\ce{{\cal E}}

\def\calL{{\cal L}}

\def\R{{\rm R}}

\def\half{\frac{1}{2}}

\def\widebar{\overline}
\def\ul{\underline}

\def\sb{\bar{\sigma}}

\def\jb{\widebar{\psi}}

\def\Jb{\widebar{\Psi}}

\def\da{{\dot \alpha}}
\def\db{{\dot \beta}}
\def\dg{{\dot \gamma}}

\def\et{{\tilde \e}}
\def\Ct{{\widetilde C}}

\catcode`\@=11

\def\ppnumber#1{\gdef\@ppnumber{#1}}

\def\@maketitle{%
  \newpage
  \null
  \begin{flushright}
    \@ppnumber
  \end{flushright}
  \vskip 2em%
  \begin{center}%
  \let \footnote \thanks
    {\LARGE \@title \par}%
    \vskip 1.5em%
    {\large
      \lineskip .5em%
      \begin{tabular}[t]{c}%
        \@author
      \end{tabular}\par}%
    \vskip 1em%
    {\large \@date}%
  \end{center}%
  \par
  \vskip 1.5em}
\catcode`\@=12

\begin{document}

\title{{\bf The Superalgebraic Approach to Supergravity}}

\author{
C.R. Preitschopf\\
Institut f\"ur Physik, Humboldt-Universit\"at zu Berlin\\
Invalidenstra\ss e 110, D-10115 Berlin, Germany\\
\and
M.A. Vasiliev\\
I.E. Tamm Theoretical Department, Lebedev Physical Institute\\
Leninsky Prospect 53, 117924 Moscow, Russia}

\ppnumber{FIAN/TD/15-98 \\ HUB EP-98/29 \\ hep-th/9805127}

\date{}
 
\maketitle

\begin{abstract}
We formulate classical actions for N=1 supergravity in D=(1,3)
as a gauge theory of $OSp(1|4)$. One may choose the action
such that it does not include a cosmological term.
\end{abstract}

\section{Introduction 
}

The theory of supergravity \cite{deserzumino,ffn} was discovered more
than 20 years ago, but in spite of determined efforts we do not
understand some very basic properties of this theory. There is, as of
now, no manifestly supersymmetric action for the most interesting
cases, namely in $D=10$ and $D=11$. It seems that ordinary superspace
becomes exceedingly difficult to handle beyond $N=1$ in $D=4$. The
introduction of harmonic variables \cite{harmony1,harmony2} helps
significantly, but they are introduced more as a (very clever) trick
than from first principles. A further mystery arises from
nonperturbative phenomena in string theory: the supersymmetry algebra
includes tensor charges which point to $OSp(1|32)$, but the relation
of this symmetry group to $D=11$ supergravity \cite{cjs}, though
suspected from the beginning, remains unclear.

It seems that we are missing some essential ingredients in the
formulation of supergravity, which probably are not strictly necessary
in the case $N=1$, $D=4$, but which become indipensable in higher
dimensions or for higher $N$. In this paper we take a first step in our
search for those ingredients by (partially) disentangling gauge
symmetries and reparametrizations.

This was first done by MacDowell and Mansouri for gravity and
supergravity \cite{macdowell}. The gauged superalgebra approach to
supergravities they used is developed and carefully explained in
\cite{pvn}. A formulation in terms of compensating fields was given
by Chamseddine \cite{chamseddine} for gravity coupled to Spins 3/2, 1
and 1/2. The gravity case was examined in detail by Stelle and West
\cite{stellewest}.  We present supergravity as a partially
compensated gauge theory of $OSp(1|4)$.

\section{
Gravity as a gauge theory}

In order to describe gravity as a gauge theory of $SO(2,3)=Sp(4,R)$
(in the anti-de Sitter case) or $SO(1,4)$ (in the de Sitter case),
one introduces a tangent space metric
$\eta^{MN}=(-1,1,1,1,\mp 1)$ and a connection 1-form
$\o^{MN}$, with $N=0,1,2,3,5$. The connection should decompose
as the usual fourdimensional Lorentz connection $\o^{mn}$ and the
vierbein $e^{m}=\r \o^{m5}$. We may perform this split in a
gauge-covariant way at the cost of introducing a compensator field
$U^{M}$ that satisfies $U^{M} U_{M} = \mp \r^2$.
Then the role of the vierbein is played by the frame field
\be
E^M = DU^M \ ,
\ee
which has the property $E^M U_M=0$. The covariant Lorentz-connection
\be
\o^{MN}_\calL = \o^{MN} \mp \frac{1}{\r^2} \left( U^M E^N - E^M U^N \right)\ ,
\ee
is defined by $D_\calL U^M = d U^M + \o^{MN}_\calL U_N = 0$.
$U^M$ describes locally a fourdimensional subspace of the fivedimensional
tangent space we started with.
The de Sitter curvature
\be
\half R^{MN} = 
\half dx^{\ul m}dx^{\ul n} R_{\ul m \ul n}{}^{MN}
 = d \o^{MN} + \o^{M}{}_K \o^{KN}
\ee
decomposes as follows:
\be
\half R^{MN} = \half R^{MN}_\calL
                \pm \frac{1}{\r^2} \left( U^M T^N - T^M U^N \right)
                \pm \frac{1}{\r^2} E^M E^N \ ,
\ee
with
$\half R^{MN}_\calL = d \o^{MN}_\calL  + \o_\calL ^{M}{}_K \o^{KN}_\calL$
and $T^M = DE^M = \half R^{MN} U_N$. The action
\bea
S & = & \mp \frac{1}{64 \pi \r} \int_{M_4}  \epsilon_{N_1 \ldots N_5}
U^{N_1} R^{N_2 N_3} R^{N_4 N_5} \nn \\
 & = &  \mp 
\frac{1}{64 \pi \r} \int_{M_4}  \epsilon_{N_1 \ldots N_5} U^{N_1}
\left(  R_\calL^{N_2 N_3} R_\calL^{N_4 N_5} 
        \pm \frac{4}{\r^2} E^{N_2} E^{N_3} R_\calL^{N_4 N_5}
        +  \frac{4}{\r^4} E^{N_2} E^{N_3} E^{N_4} E^{N_5} \right)\nn \\
\label{boseaction}
\eea
($\epsilon^{01235} = 1$)
is manifestly reparametrization invariant and de Sitter gauge invariant.
Upon gauge fixing $U^M = \mp \r \de^M_5$ we obtain
\be
S  = \mp  2\pi {\chi}(M_4) - \frac{1}{2 \k^2} \int_{M_4} |e| R(e,\o) 
\pm \frac{6 \pi}{\k^4} \int_{M_4} |e|  
\ee
with $R= e_m{}^{\ul m} e_n{}^{\ul n} R_{\ul m\ul n}{}^{nm}$
and $\k^2 = 2 \pi \r^2$, i.e. the usual Einstein action with a cosmological
constant and a topological term. The presence of such terms is due to the
simple choice of $S$ above, and has nothing to do with the fact that we
formulated a gauge theory of the (anti-) de Sitter group. We may write a
slightly more complex action that reproduces precisely Einstein gravity:
\be
S_\ce  = - \frac{1}{8 \k^2 \r} \int_{M_4}  
\epsilon_{N_1 \ldots N_5} U^{N_1} E^{N_1} E^{N_2} R_\calL^{N_4 N_5} 
=  - \frac{1}{2 \k^2} \int_{M_4} |e| R(e,\o) \ ,
\label{einsteinaction}
\ee
which is again reparametrization and gauge covariant.
We learn that the vacuum algebra, i.e. the symmetry algebra of
(anti-) de Sitter or Minkowski space has little to do with the gauge
algebra: we may choose the Poincare or the (anti-) de Sitter algebra
for either one, and independently. The main advantage of choosing
(\ref{boseaction}) or (\ref{einsteinaction}) is that the vacuum
solutions $R^{MN}=0$ resp. $R_\calL{}_{mn}{}^{np}=0$ can be read
off the action immediately.

\section{
$Sp(4,R)$}

In order to formulate supergravity as a gauge theory it will prove
useful to convert the vector notation of the previous section to
a spinor one. We define the fundamental representation of
$Sp(4,R)\sim O(2,3)$
in terms of an $Sl(2,C)$-spinor as
\be
\Psi^a = \left(\Psi^\a, \Jb^\da  \right)\ ,
\ee
with $\left(\Psi^\a\right)^\ast = \Jb^\da$. The invariant tensor
$ C_{ab} = ( \e_{\a\b}, \e_{\da\db}) $ and its inverse
$ \Ct^{ab} = ( \et^{\a\b}, \et^{\da\db}) $ may be used to
raise and lower spinor indices. The $Sp(4,R)$-generators $J_{ab}$
are symmetric bispinors.
The covariant derivative $D\Psi^a = d\Psi^a + \o^a{}_b \Psi^b$
leads to the curvatures $ \half R^{ab} = d \o^{ab} + \o^{a}_{c} \o^{cb}$,
which decompose as follows: 
\bea
\R^\a{}_\b & = & \frac{1}{4} \ \s_{mn}{}^\a{}_\b\  R^{mn} \quad ; \quad
\widebar\R^\da{}_\db = \frac{1}{4} \ \sb_{mn}{}^\da{}_\db\  R^{mn} \nn\\
\R^\a{}_\db & = & \half \s_{m}{}^\a{}_\db\  R^{m5} \quad ; \quad
\R^\da{}_\b = \half  \sb_{m}{}^\da{}_\b\  R^{m5} \ .
\label{curvdecomp}
\eea
The vector field $U^M$ now appears as an antisymmetric traceless bispinor:
\be
U^{[ab]} \ = \ i \left(
\begin{array}[c]{cc}
 U^5 \e^{\a\b} &  U_m \s^m{}^{\a\db} \\
 -U_m \sb^m{}^{\da\b} & -U^5 \e^{\da\db} \\
\end{array}
\right) \ .
\ee
After gauge fixing we obtain the action (\ref{boseaction}) in the form given
by MacDowell and Mansouri \cite{macdowell}:
\be
S = \frac{i \r^2}{8 \k^2} \int_{M_4}
 \left(R^{\a}{}_{\b} R^{\b}{}_{\a} - R^{\da}{}_{\db} R^{\db}{}_{\da} \right )\ .
\ee

\section{
$OSp(1|4)$ and Supergravity}

We now upgrade the gauge algebra to $OSp(1|4)$, with fundamental
representation $\Psi^A = (\Psi^a, \Psi^j ) = (\Psi^A)^\ast$, grading
\be
(-)^A \ = \ \Bigg\{
  \begin{array}[c]{cccc}
  +1 &{\rm for} &A \in \{\a, \da\} \quad; & \a, \da \in \{1,2\}\\
  -1 &{\rm for} &A = j \quad; &j \in \{1\} \\
  \end{array}
\ee
and invariant tensors
\bea
C_{[AB)} = -(-)^A C_{BA} = \left( C_{ab}, i \de_{ij} \right) & ; &
\Ct^{[AB)} = -(-)^A \Ct^{BA} = \left( \Ct^{ab}, -i \de^{ij} \right)
\nn\\
C_{AB} \Ct^{BC} = \Ct^{CB} C_{BA} & = & \de_A^C = (\de_a^c , \de_i^k)\ ,
\eea
which raise and lower indices as follows:
\be
\Psi_A \ = \ C_{AB} \Psi^B \quad ; \quad
\Psi^A \ = \ \Ct^{AB} \Psi_B \ .
\ee
The standard index contraction is
\be
\J_A \Phi^A = -(-)^A \J^A \Phi_A \quad ; \quad
\left(\J_A \Phi^A\right)^\ast =  \Phi^A \J_A \ ,
\ee
and upon introducing the graded symmetric
gauge connection $\o^{(AB]}$ we arrive at curvatures $R^{AB}$,
which now contain fermionic (gravitino) curvatures $R^{aj}$ in addition
to those listed in (\ref{curvdecomp}).
The compensator field $U^{[AB)}$ is graded antisymmetric and traceless,
and we require it to satisfy the conditions \cite{chamseddine}
\be
U_{A}{}^{A} = 0  \quad ; \quad
U_{AB} U^{BA} = 4 \r^2  \quad ; \quad U_{AB} U^B{}_C U^{CA} = 0 \ .
\label{ucon}
\ee
The action
\be
S = \frac{\r}{4 \k^2}
\int_{M_4} U^A{}_B R^B{}_C \left(
\delta^C{}_E + \frac{1}{2 \r^2} U^C{}_D U^D{}_E \right)
R^E{}_A (-)^{A}
\label{superaction}
\ee
is then manifestly reparametrization invariant (since it is a 4-form),
and $OSp(1|4)$-gauge invariant. In addition, it possesses a local
supersymmetry which stays hidden because we are not working in
superspace.
We gauge fix $U^{AB}$ by the local $OSp(1|4)$ as follows:
\be
U^A{}_B = \left(
 \begin{array}[c]{ccc}
   i \r \de^\a{}_\b   &                  0   & 0 \\
   0                  & -i \r \de^\da{}_\db  & 0 \\
   0                  & 0                    & 0 \\
 \end{array}
\right) \ ,
\label{gaugefix}
\ee
and (\ref{superaction}) takes the form
\be
S = \frac{i \r^2}{8 \k^2} \int_{M_4}
 \left(R^{\a}{}_{\b} R^{\b}{}_{\a} + 2 R^{\a}{}_j R^j{}_{\a}
- R^{\da}{}_{\db} R^{\db}{}_{\da} - 2 R^{\da}{}_j R^j{}_{\da}\right) \ ,
\ee
which may be rewritten as
\bea
S & = &
- \frac{\r^2}{32 \k^2} \int_{M_4} d
    \left[ \e_{mnpq} \big( \o^{mn} d \o^{pq} +
                           \frac{2}{3}\o^{mn} \o^{pl} \o_l{}^{q} \big)
     + \frac{16 \k^2}{\r} \j^\a  D^{\cal L}  \j_\a
     - \frac{16 \k^2}{\r} \jb^\da  D^{\cal L}  \jb_\da
    \right] \nn\\
 && - \frac{1}{2 \k^2} \int_{M_4} |e| R(e,\o) \nn\\
 && + \frac{1}{2} \int_{M_4} |e| \e^{mnpq} \left(
           \jb_m{}^\da \sb_{n}{}_\da{}^\b D^{\cal L}_p \j_{q\b} -
           \j_m{}^\a \s_{n}{}_\a{}^\db D^{\cal L}_p \jb_{q\db} \right) \nn\\
 && + \int_{M_4} |e| \left( \frac{3}{\k^2 \r^2} - \frac{i}{2\r}
       \Big[ \j_{m\a} \s^{mn}{}^\a{}_\b \j_n{}^\b +
       \jb_{m\da} \sb^{mn}{}^\da{}_\db \jb_n{}^\db \Big] \right) \ ,
\label{superdecomp}
\eea
where $D^{\cal L}_m$ contains only the Lorentz connection. It is
well known that (\ref{superdecomp}) is supersymmetric.

\section{
Supersymmetry}

In order to make this symmetry transparent, we compute the variation
of the above action under infinitesimal variations of the fermionic
compensators:
\be
\de S \propto \int_{M_4} R^\a{}_\db \left(
\de U^\db{}_j R^j{}_\a + R^\db{}_j \de U^j{}_\a \right)
\label{variu}
\ee
and compare it with the effect of arbitrary infinitesimal
variations of the Lorentz connection:
\be
\de S \propto \int_{M_4} R^\a{}_\db \o^\db{}_\g \de \o^\g{}_\a
-  R^\da{}_\b \o^\b{}_\dg \de \o^\dg{}_\da \ .
\label{vario}
\ee
we may either determine $\de \o^\a{}_\b, \de \o^\da{}_\db$ such that
the sum of both variations vanishes (1st order formalism) or we
may solve the algebraic equations of motion for
$\o^\a{}_\b, \o^\da{}_\db$, i.e. set $R^\a{}_\db=0$ (1.5 order formalism).
In either case we have shown that there is a hidden fermionic
symmetry, which turns out to be supersymmetry:
\bea
\de \o^\a{}_\db & = & \o^\a{}_j \l^j{}_\db - \l^\a{}_j \o^j{}_\db \nn\\
\de \o^\a{}_j & = & D^{\cal L} \l^\a{}_j + \o^\a{}_\db \l^\db{}_j \ .
\eea
In this picture supersymmetry looks like a gauge symmetry, because it is
inherited from a bona fide gauge theory of a graded Lie algebra. The
hidden aspect of this symmetry is, that in the sense
expressed by (\ref{variu}) and (\ref{vario}), the action
is in fact independent of the fermionic fields $U^{aj}$, and hence the
fermionic gauge symmetry is true and uncompensated.

\section{\label{proj}
Projectors}

The compensators $U^{AB}$ allow us to distinguish
an $Sl(2,C)$ subgroup within $OSp(1|4)$ in a gauge covariant way.
This is most clearly seen by the construction of projection
operators:
\bea
\Pi{}^{(a)}{}_{(b)} & = & - \frac{1}{\r^2}\ U^A{}_C U^C{}_B =
              \left( \de^\a{}_\b , \de^\da{}_\db , 0 \right) \nn\\
\Pi{}^{(\a)}{}_{(\b)} & = & \half  \left(- \frac{i}{\r}  U^A{}_B
              - \frac{1}{\r^2} U^A{}_C U^C{}_B \right) =
              \left( \de^\a{}_\b ,  0 , 0 \right) \nn\\
\Pi{}^{(\da)}{}_{(\db)} & = & \half  \left( \frac{i}{\r}  U^A{}_B
              - \frac{1}{\r^2} U^A{}_C U^C{}_B \right) =
              \left(  0 , \de^\da{}_\db ,  0 \right) \nn\\
\Pi{}^{(i)}{}_{(j)} & = & \de^A{}_B  + \frac{1}{\r^2} U^A{}_C U^C{}_B =
              \left(  0  ,  0 , \de^i{}_j \right)  \ .
\eea
The last equality in each line holds of course only after fixing the
gauge (\ref{gaugefix}). Proving the projection property requires some
computation, but relies only on the properties (\ref{ucon}) of the 
compensator field, which imply among other things 
$U^a{}_a = \frac{1}{\rho^2} U^a{}_b  U^b{}_j  U^j{}_a$
and as a consequence $U^A{}_B U^B{}_C  U^C{}_D = -\r^2 U^A{}_D$.
We use the notation $\J^{(\a)}= \Pi{}^{(\a)}{}_{(\b)}\J^{B}$, and
may then formulate (\ref{superaction}) as
\be
S = \frac{i \r^2}{8 \k^2} \int_{M_4}
 \left( R^{(\a)}{}_{(\b)} R^{(\b)}{}_{(\a)} + 2 R^{(\a)}{}_{(j)} R^{(j)}{}_{(\a)}
- R^{(\da)}{}_{(\db)} R^{(\db)}{}_{(\da)} - 2 R^{(\da)}{}_{(j)}
R^{(j)}{}_{(\da)}\right) \ ,
\ee
which is the $OSp(1|4)$-invariant form of the MacDowell-Mansouri action.
Once one knows the projectors
it is also possible to perform the $Sl(2,C)$-decomposition covariantly.
We set for simplicity $\r=1$ and start with
\bea
\o^A{}_B & = & \o_\calL{}^A{}_B
+ U^A{}_C E^C{}_D (1+U^2)^D{}_B -  (1+U^2)^A{}_C E^C{}_D U^D{}_B \nn\\
&& - \frac{1}{4}\left(
             U^A{}_C E^C{}_D (U^2)^D{}_B - (U^2)^A{}_C E^C{}_D U^D{}_B
             \right)  \ ,
\eea
where the Lorentz connection $\o_\calL{}^A{}_B$ is defined by
$D^\calL U^{AB} =0$, and $E^A{}_B = DU^A{}_B$.
Then (\ref{superaction}) is decomposed as follows:
\bea
S & = &
\frac{1}{8 \k^2}
\int_{M_4} (-)^A U^A{}_B \Big[ R_\calL R_\calL \Big]^B{}_A
+ 8 d \left( (-)^A U^A{}_B E^B{}_C D^\calL  E^C{}_A \right) \nn\\
&&+ 4 (-)^A U^A{}_B \Big[ E (1+U^2) R_\calL E -
                      E (1+U^2) E U^2 E (1+U^2) E  \Big]^B{}_A \nn\\
&&- (-)^A U^A{}_B
  \Big[ E U^2 E R_\calL - 8 E U E D^\calL E\Big]^B{}_A \nn\\
&& + (-)^A U^A{}_B 
  \Big[ \frac{1}{4} E U E U E U E U + 4 E U E U E (1+U^2) E \Big]^B{}_A \ .
\nn\\
\label{ospcov}
\eea
The first line of (\ref{ospcov}) is topological, the second line
vanishes upon gauge fixing, since the index $j$ can take only one value,
the third line yields the kinetic terms for gravitons and gravitinos,
and the last line describes the supersymmetric cosmological constant.

\section{
The Cosmological Term}

The decomposition of the action (\ref{superaction}) in the form
(\ref{superdecomp}) shows that we have obtained a supersymmetric
cosmological term. One may be tempted to regard this as an inherent
feature of any $OSp(1|4)$ gauge theory of gravity. (\ref{ospcov})
implies that this is not so. We easily extract pure supergravity
without cosmological constant:
\bea
S_{{\cal E}+3/2} & = & \frac{1}{8 \r \k^2}
 \int_{M_4} (-)^A U^A{}_B \Big( 4 DU^B{}_{(c)} DU^{(c)}{}_D R^D{}_A
- 3 DU^B{}_{(c)} DU^{(c)}{}_{(d)} R^{(d)}{}_A \nn\\ &&
- \frac{2}{\r^2} DU^B{}_C DU^C{}_D DU^D{}_E DU^E{}_A
+ \frac{3}{2 \r^2} DU^B{}_{(c)} DU^{(c)}{}_{(d)} DU^{(d)}{}_{(e)} DU^{(e)}{}_A
\Big) \nn\\
 & = &
- \frac{1}{8 \k^2\r^3} \int_{M_4} (-)^A U^A{}_B
  \Big[ E U^2 E R_\calL - 8 E U E D^\calL E\Big]^B{}_A \nn\\
 & = &
 - \frac{1}{2 \k^2} \int_{M_4} |e| R(e,\o)
 + \frac{1}{2} \int_{M_4} |e| \e^{mnpq} \left(
           \jb_m{}^\da \sb_{n}{}_\da{}^\b D^{\cal L}_p \j_{q\b} -
           \j_m{}^\a \s_{n}{}_\a{}^\db D^{\cal L}_p \jb_{q\db} \right) \ .
\nn\\
\label{supereinstein}
\eea
Admittedly the first line of (\ref{supereinstein}) does not look
particularly elegant. We understand it only by rewriting it in the
form of line two. Without mentioning compensators Chamseddine and West
wrote this action quite early in the development of
supergravity \cite{chamwest}. It proves our point that the
gauge algebra is independent of the vacuum algebra, i.e. the algebra
of isometries of the vacuum state.

\section{
Conclusions}

We have presented both gravity and supergravity as partially
compensated gauge theories. The results of section \ref{proj} make it
easy to formulate any Lorentz-covariant theory in terms of $OSp(1|4)$,
without the need for a group contraction. The significance of this
gauge group lies then in the simplicity of (\ref{superaction}) and the
hidden supersymmetry.

We believe that one can extend this anaysis to include additional
fermionic as well as bosonic coordinates. This should lead to a
natural and simple form of supergravity in superspace.

\section*{Acknowledgments}
Part of this work was done at Utrecht. We thank B. de Wit and the
theory group for hospitality and a productive work environment. This
work was supported in part by the following grants: DFG 436 RUS 113,
EC network ERBFMRXCT96-0045, RFBR 02-17314, INTAS 96-0538 and 
INTAS 93-0023.

\end{document}